\documentclass[acmsmall]{acmart}

\usepackage{soul}
\usepackage{verbatim}

\makeatletter
\AtBeginDocument{\let\hl\@firstofone}
\makeatother

\AtBeginDocument{%
  \providecommand\BibTeX{{%
    \normalfont B\kern-0.5em{\scshape i\kern-0.25em b}\kern-0.8em\TeX}}}

\setcopyright{rightsretained}
\acmJournal{PACMHCI}
\acmYear{2021} \acmVolume{5} \acmNumber{CSCW1} \acmArticle{146} \acmMonth{4} \acmPrice{} \acmDOI{10.1145/3449220}

\begin{document}

\title{Characterizing the Online Learning Landscape: What and How People Learn Online}

\author{Sean Kross}
\email{seankross@ucsd.edu}
\orcid{0000-0001-5215-0316}
\affiliation{%
  \institution{University of California San Diego}
  \country{USA}
}

\author{Eszter Hargittai}
\affiliation{\institution{University of Zurich}\country{Switzerland}}
\email{pubs@webuse.org}

\author{Elissa M.\ Redmiles}
\affiliation{\institution{Max Planck Institute for Software Systems}\country{Germany}}
\email{eredmiles@gmail.com}


\begin{abstract}
Hundreds of millions of people learn something new online every day. Simultaneously, the study of online education has blossomed within the human computer interaction community, with new systems, experiments, and observations creating and exploring previously undiscovered online learning environments. In this study we endeavor to characterize this entire landscape of online learning experiences using a national survey of 2260 US adults who are balanced to match the demographics of the U.S. We examine the online learning resources that they consult, and we analyze the subjects that they pursue using those resources. Furthermore, we compare both formal and informal online learning experiences on a larger scale than has ever been done before, to our knowledge, to better understand which subjects people are seeking for intensive study. We find that there is a core set of online learning experiences that are central to other experiences and these are shared among the majority of people who learn online. We conclude by showing how looking outside of these core online learning experiences can reveal opportunities for innovation in online education.
\end{abstract}

\ccsdesc[500]{Applied computing~E-learning}
\ccsdesc[500]{Human-centered computing~Human computer interaction (HCI)}

\keywords{Online learning; Adult learning; Survey; MOOCs; YouTube}

\maketitle

\section{Introduction}

The subject of how people learn has been fascinating researchers since long 
before the invention of the 
internet ~\cite{bradford-1958,mower-1960,rogers-1957,stengel-1939}. The 
innovation of online learning has opened the door for more people to learn, 
and thus, the door for more research into what people are learning about and 
how they are doing so.

One of the most promising aspects of the internet since its early days has 
been the opportunities it offers for widespread participation thanks to the 
myriad of resources it makes freely available ~\cite{Benkler-2006}. While 
much scholarly work has investigated what this means for the development of 
resources like Wikipedia and other wikis (e.g., Shaw \& Hill, 2014), 
researchers have paid much less attention to how people use diverse online 
resources for educational purposes ~\cite{shaw-2014}. The internet offers a 
plethora of formal and informal online resources available for free or at 
significantly lower cost compared to traditional, offline educational 
opportunities, and has thus widened the availability of educational 
resources. Massive open online courses (MOOCs), video tutorials, how-to articles, 
online discussion groups, among others, can help people gain new skills with 
the potential to improve their job prospects, social mobility, 
and personal welfare~\cite{hadavand2018can}.

Prior work has considered how a specific subset of people learn (e.g., 
young adults), how people learn a specific subject, or how people use a 
particular resource for 
learning ~\cite{shorey-2020,kross2020democratization,Narayan-2017,Torrey-2009}. Yet, little 
is known about the \emph{full ecosystem} of what adults learn about online and what 
educational resources they use to do so. These can include both formal online 
courses (e.g., Coursera, those offered by local universities) and less formal 
resources (e.g., YouTube videos, online discussion groups). No prior work, to 
our knowledge, has addressed this full ecosystem by looking at more than one 
subject or one resource at a time across all online learners. An important 
novel contribution for this paper is that it considers the \emph{combination} of 
subjects and resources (e.g., an online university course for learning 
history, YouTube for learning math) that may generalize to many learners. 

In this work, we explore three research questions: 

\textbf{RQ1)} What do people learn about online?

\textbf{RQ2)} How do people learn online (i.e., what learning resources do they use)?

\textbf{RQ3)} What are the core online learning experiences - which subject through what resource - that are common across the majority of online learners?

To answer these research questions, we conducted an online survey of 2,260 adults age 18 and over in the U.S. To improve the generalizability of our findings, our survey sample was demographically balanced to match U.S. Census statistics on gender, age, and education. We draw on frameworks and findings of past literature on online learning ~\cite{Piety-2014,Ferguson-2012,kross-2018} to analyze our findings critically and characterize the online learning ecosystem.

We find that the vast majority (93\%) of those we surveyed had learned something online. At a high level, we find that online learners' interests span a very wide set of subjects (with the median subject being learned by 26\% of respondents) and resources (with the median resource being used by 62.5\% of respondents). Further, examining online learning experiences (pairs of subjects and resources), we find 12 core online learning experiences shared by the plurality of online learners. The most common of these core online learning experiences, which was reported by over half of our respondents, was learning how to do something yourself (DIY) using YouTube. 

The findings from this work provide insight into the ecosystem and experience of online learning; suggest a combination of factors that may drive learners' choices and affect our progress toward a democratization of online learning; and identify directions for future work on improving the design and targeting of online learning technologies to draw in new learners and onboard new subjects.
\section{Related Work}
\label{sec:related}

\subsection{Frameworks for Quantitative Analyses of Online Education} 

This work is modeled after surveys of participants in sociotechnical systems 
that are typical throughout the Computer Supported Cooperative Work and Human 
Computer Interaction communities. Our work is inspired specifically by three 
analysis frames posited by prior work.

First, we draw from the work of Piety et al., who present a framework for the 
educational data sciences, which delineates studies by (1) the educational 
stage of the learners studied, and (2) the unit of analysis through which the 
learning is analyzed~\cite{Piety-2014}. We use this framework to design our 
study: we choose to study (1) adult learners and (2) our unit of analysis is the individual with a focus on the subjects and resources they consult online for learning.

Piety and colleagues additionally explain that a challenge of educational 
data is that they are often limited to specific types of online learning 
resources. In order to address this limitation and maximize the possible set 
of resources available for us to analyze, we sample a large and 
demographically diverse set of respondents in our study. By collecting 
survey responses from people whose demographics are proportional to the 
overall US population on multiple demographic axes, we aim to capture a 
set of learners who use a broad set of resources. 

In addition to the Piety et al. framework for educational data science 
studies, we draw from the analysis frame of Ferguson and Shum. Ferguson 
and Shum argue for always using networked analysis to understand online 
learning~\cite{Ferguson-2012}. Their rationale for this methodological 
approach is that online learners often consult multiple types of online 
learning resources when trying to understand a subject, and therefore the 
relationship between the different types of resources should be understood 
better. They point to the proliferation of Open Educational Resources (OERs) 
that learners are able to consult online as a basis for the need to 
characterize how the uses of these resources intersect. Following their 
suggestion, we use clustering analysis to understand our data and draw from 
these analyses to identify opportunities for developing new online learning 
experiences. 

Finally, our analysis has been influenced by 
the work of Kross and Guo, who provide a systematic 
review of the literature on how different types of online learning resources 
connect students to each other to create new interactions that shape 
students' learning trajectories~\cite{kross-2018}. They highlight how 
software systems can be purpose-built to enhance learning experiences for 
certain subjects. Applying this purpose-driven lens to our results, 
we will discuss the implications of our empirical analysis on the structure of 
the online learning communities described in their study.

\subsection{Surveys of Online Educational Experiences} 

Prior work has investigated various aspects of learners' online educational 
experiences. Kizilcec and Schneider developed and deployed the Online Learning 
Enrollment Intentions scale as an instrument for understanding the motivations 
for learners to seek online education ~\cite{Kizilcec-2015}. They found that 
online learners often have social goals when taking a course, which contrasted 
to the typical course design built upon assignments meant to be completed 
alone. They further found that the design of online educational resources 
should incorporate an awareness that learners are using multiple resources, 
and that online instructors should direct learners toward other resources. 
In our study we extend these findings beyond the domain of online courses to 
quantify the extent to which ten types of online learning resources are used 
alongside other types of resources.

Additionally, Swanson and Walker surveyed young adults between the ages of 
18 and 25 to gain insight into their usage of several different digital 
technologies for academic and recreational purposes ~\cite{swanson-2015}. 
They found that young adults spend a majority of out-of-classroom 
academically-focused time using technology. Their results inform how 
different types of online learning resources could be deployed depending on 
how young adults are using different devices. Our study builds upon their work 
by analyzing how different types of online learning resources are used, which 
may be influenced by what subjects can be studied effectively given the device 
that is being used. Both of these studies aim to understand how technology 
can be harnessed to meet learners' expectations in terms of where they can 
find online educational resources and how they can be supported to have 
fulfilling learning experiences via those resources.

\subsection{Formal and Informal Online Learning Experiences} 

The online education literature includes studies of both more formal online 
educational experiences like online courses offered for college credit and 
massively open online courses that award credentials
\cite{hew-2014,pastore-2009}. The literature also includes studies of more 
casual online educational experiences that are related to the field of 
``free-choice learning,'' which includes going to a museum or reading a non-fiction
book for pleasure~\cite{falk-2007}. 
For our discussion about online learning resources we will differentiate between
\emph{formal} and \emph{informal} resources based on Rosenthal's work on 
free-choice learning about YouTube, where both \emph{formal} and \emph{informal}
resources have learning outcomes but only \emph{formal} resources have 
prescribed learning objectives and \emph{informal} learning experiences take
place ``outside a formal learning environment''~\cite{Rosenthal-2018}. 
Additionally, Wenger characterizes informal learning as requiring community participation, versus formal learning that takes place in a classroom or a structured learning environment~\cite{wenger-2002}.
In line with prior work, we refer to enrollment in an online course as use of a 
\emph{formal} 
%
%
online learning resource, while we refer to 
all other types of online learning resources as 
\emph{informal}.
%
%
We believe that 
this terminology is justified considering the time and financial commitment 
that many online courses require, and the resemblance that online courses 
have to traditional in-person learning experiences in the classroom. 

Studies of formal online learning resources greatly outnumber studies of 
informal learning resources, perhaps due to the high level of access that 
university researchers have to such learners and data generated from online 
university courses~\cite{kross-2018,kross2020democratization,Zheng-2015}. These studies 
often focus on students' level of engagement with online course materials, 
how many students complete the course compared to how many enroll, and how 
these dynamics differ from in-person educational experiences.

Here we highlight a few examples of studies about informal online learning 
that are especially salient to the goals of our study. Many studies of 
informal online learning are centered around one specific aspect of the 
educational experience, usually a specific type of online learning resource, 
or a study about how a particular subject is pursued online. In an example 
of the former, Narayan et al. developed and studied the success of an 
interactive tutorial to onboard new members of the Wikipedia 
community~\cite{Narayan-2017}. Although the results from their intervention 
did not cause participants to be any more likely to contribute to Wikipedia, 
it made participants feel more integrated into the Wikipedia community. 
Torrey et al. explored teaching and learning crafts online via a series of 
interviews focused on how members of craft communities communicate ideas 
about space and aesthetics that are not easily communicated with typical 
online resources ~\cite{Torrey-2009}. They found that curation of specific 
online resources and persistence in communicating complex ideas about their 
craft were key to their success in this informal learning community. 

Yet other studies on informal online learning focused on how people learn a 
specific subject. Exemplar studies include that by Shorey et al., who studied how 
participation in social subgroups within an online community for the Scratch 
educational programming language ~\cite{resnick-2009} leads to higher levels 
of engagement and enriching interactions that may lead to better learning 
outcomes ~\cite{shorey-2020}. These results are further supported by 
Yang et al. and Gelman et al., who studied how communities of informal 
learning can grow, and how learning trajectories can be mapped in informal 
learning settings ~\cite{Yang-2015,Gelman-2016}. 

Our study differs from past work on both formal and informal learning in that 
we (a) study a far broader set of resources and subjects, aiming to 
characterize and critically examine the full ecosystem of online learning, 
and (b) while prior work has called for the comparison of formal and 
informal resources ~\cite{schwier-2012}, our work is the first that we are 
aware of to compare these mechanisms of online learning empirically
(specifically, in terms of the subjects that people learn using differing resource types).
\section{Methods}

We conducted a national survey of American adults to examine the relationship between 
the subjects that people learn about online and the types of online learning 
resources they use to do so. Here, we provide details about the data 
collection, the statistical procedures we applied to the data, and the 
limitations of our analysis.

\subsection{Data Collection}

We contracted with the survey research firm Cint to administer the study to 
American adults in June-July, 2019. People were able to fill out the survey 
hosted on the Qualtrics platform using a computer or a mobile device. We 
included two attention-check verification questions and only those who passed 
both are included in the analyses. We received 2,260 valid responses to our 
survey.
%
%

We quota-sampled on gender, age, and education to obtain a diverse sample.
Specifically, our 
respondents are 58.6\% female and the mean age of our respondents is 41 years 
(SD=15 years). Additionally, 23.6\% of our sample has a high school diploma or 
less education, 39.8\% of our sample has some college education, while the 
remaining 36.6\% of our sample has a college degree or more education. The 
survey panel provider compensated respondents for their participation based 
on their preference of cash, gift cards, or donations to charity. All study
procedures and materials were approved by our Institutional Review Board.

\subsection{Survey Questionnaire}


To understand respondents' online learning experiences, we asked respondents 
what \textbf{subjects} they had learned about online and what types of \textbf{resources} they 
used to learn about each of those subjects\footnote{Respondents had the option 
to select none of the options, thus indicating that they had not learned 
online.}. A respondent who indicated that they used a particular type of 
resource to learn about a particular subject had what we define as an 
\textbf{\emph{online learning experience}}. 

To understand what subjects people learned online, we asked ``Which of the 
following topics have you tried to learn about through online resources?" This 
was followed by nineteen subjects, the option of ``other" and specifying 
something else, and ``none of the above." Subjects ranged from traditional 
academic subjects like history, math, and science to general welfare and 
lifestyle subjects such as: do-it-yourself (DIY)\footnote{This was phrased 
as ``how-to (e.g., around the house, cooking/baking, etc.)''}, travel/geography,
personal health/health care, makeup/fashion, and online safety, security, 
or privacy. The categories of subjects were developed through iterative 
rounds of 10 cognitive interviews with participants of varied demographics (age, gender, socioeconomic status). In these interviews we prompted participants about whether there were any answers they wanted to provide, but which were not available for them to select. These interviews are a commonly-used technique for testing survey 
questions~\cite{Presser-2004} and are not intended as research artifacts (see Section~\ref{sec:survey:validation} below for more details).

To understand how people learned online -- that is, what resources they use 
to learn -- we included a matrix question that asked about how participants learned 
each of the subjects they indicated in the prior question. Specifically, 
respondents could select multiple options from a list of ten resources: 
watched a video (e.g., YouTube), read Wikipedia, took an online course, 
read an informational article, read a how-to-guide, used an interactive 
tutorial, used a practice exam website, used materials from a course, 
read answers to questions on an online discussion community or forum, 
and asked questions in an online discussion community or forum. 

Other than the first three of these response options, the others included 
at least two examples (see below) in parentheses to help respondents 
understand our categories. Informational articles include articles on 
non-Wikipedia websites, such as privately-run wikis like Wikia, online 
references like dictionaries and thesauri, and research-based resources 
like museum websites. How-to guides include websites that help people 
complete projects around their home as well as step-by-step guides for 
understanding mathematics like BetterExplained.com.  Examples of interactive 
tutorials on the web we mentioned include platforms like Khan Academy, 
Codecademy, and Duolingo. Practice exam sites like Kaplan and Magoosh focus 
on preparation for standardized exams, while course materials refer to open 
repositories of slides and other artifacts from courses usually taught in 
person, like the MIT OpenCourseWare project. Finally, discussion communities 
and forums include social-media forums like Facebook Groups and Reddit, in 
addition to forums designed for professional communities like StackOverflow. 
\hl{Given our interest in comparing informal online learning to formal online learning, which we understood as the online courses, those who indicated taking online courses were then presented with an additional question: ``What kinds of organizations have provided the online course(s) that you have enrolled in? (Check all that apply)'' The answer options were: Online course at a local university; Online course at a university elsewhere; Massive open online course (Coursera, FutureLearn, edX, Udacity, Udemy);  Other online course (e.g. The Great Courses, Lynda.com).}

Overall, respondents could report up to 190 possible online learning 
experiences (pairs of subjects and resources such as learning math on 
YouTube or learning languages using an interactive tutorial), since we 
asked about how respondents learned up to 19 subjects using up to 10 online 
learning resources.\footnote{The parts of the survey questionnaire that were analyzed in this work can be found at \url{https://doi.org/10.5281/zenodo.4088916
}.}

\subsection{Survey Validation and Pretesting} 
\label{sec:survey:validation}

During the in-person cognitive interviews, we asked interviewees to think 
aloud as they were filling out the survey, and they were encouraged to share 
their thoughts if they believed that they were missing answer choices or if 
they were confused about the wording of a question. Based on the feedback 
from these interviews we rewrote and rearranged our survey questions to 
maximize their clarity and to ensure the completeness of the answer choices 
we offered. Once our survey was deployed, less than one percent (0.88\%) of 
respondents indicated that their desired answer choice was not available for 
any of our questions (they selected ``Other" in response to the question), 
confirming the completeness of the choices we provided.
 
\subsection{Data Analysis}
A respondent is considered to have learned online if they learned about at 
least one subject using at least one resource. 

To investigate what subjects people learn online and what resources they use 
to learn them (RQ1 and RQ2), we use descriptive data analysis including 
hypothesis tests and $\chi^{2}$ proportions tests as appropriate. Additionally, 
as reviewed in 
Section~\ref{sec:related},
prior work has considered formal 
and informal learning resources independently, but has not examined 
differences in what these resources are used to learn about. To fill this 
gap, we make subject-by-subject comparisons between online courses and all other 
types of online educational resources combined. We used $\chi^{2}$ proportions 
tests to make inferences about whether online courses are used more, less, or 
equally as often to learn about certain subjects compared to other types of 
resources. To reduce the Type I error rate we applied the Bonferroni-Holm 
correction to the resulting p-values to account for multiple hypothesis testing.

To answer RQ3 we examined commonalities in subjects and resources by using k-modes 
clustering ~\cite{huang-1998} to organize respondents into clusters based on 
the subjects they reported learning about online and, separately, the types 
of resources they reported using. K-modes clustering is an extension of 
k-means clustering for categorical data. While k-means minimizes the distance 
between the center of each cluster and points belonging to that cluster in 
euclidean space, k-modes aims to maximize the similarity in categories shared 
between observations (respondents in our case) within a cluster. The number of 
clusters we selected for clustering respondents according to shared types of 
resources and subjects was chosen using the silhouette method 
~\cite{Rousseeuw-1987}, where we calculated the average silhouette width 
for $k$ equal to 3 through 10, and then selected the value of $k$ that 
maximized the average silhouette width. This method for selecting $k$ ensures 
that the observations being clustered are the most similar to the cluster 
they are in and that they are the most dissimilar from clusters from which they
are excluded.

\subsection{Limitations}
Respondents were asked to self-report their online learning experiences, 
therefore this study exhibits many of the same limitations as other 
self-report studies. These limitations include over- and under-reporting, 
which occur when survey results reflect an over- or under-estimate of the 
rate or abundance of an experience. This discrepancy can be caused by 
respondents misinterpreting survey questions, or it can be induced by one 
of several biases, including desirability bias (when respondents give 
socially desirable, instead of honest answers), and recall bias (when 
respondents incorrectly remember an experience). To alleviate the potential 
for these inconsistencies, we revised the survey questions iteratively 
through a series of interviews, followed by pre-testing the survey to ensure 
that respondents thoroughly answered all questions and interpreted them the 
way we had intended.

Additionally, our survey was conducted online. As our focus is to study online learning, collecting the data online is appropriate. Cint, the firm through which we collected our data, uses a double opt-in procedure to recruit respondents to its panel. Research in the past decade has established that there are ``few or no significant differences between traditional modes [of survey administration] and opt-in online survey approaches'' for research such as that presented here, which is intended to improve our understanding of human behavior and experiences~\cite{ansolabehere2014does,prepared2010research}.

The results of this survey only reflect the usage of certain types of online 
learning resources, and the subjects that respondents were interested in 
learning about. Therefore we cannot make any claims about how often these 
resources were consulted, or the duration of time that respondents spent 
pursuing particular subjects. Our findings are accompanied by a number of 
theoretical explanations that include suggestions for how online educational 
experiences could be improved and how they could be studied in the future. 
This is not an experimental study and we have data from one point in time 
so it is inappropriate to interpret any of the relationships that we 
present as causally linked.

\hl{Finally, it is important to clarify that the conception of ``learning'' 
presented in this paper is based in the computer and cognitive sciences, 
and therefore represents only one perspective among many. For example, much 
of the prior work that inspired the design of this study is related
or adjacent to studies of ``online learning.'' This includes both empirical 
studies of what kinds of activities benefit specific learning 
outcomes}~\cite{koed-15}, \hl{and theoretical work that centers 
``instructional events'' and ``learning events'' within greater learning 
frameworks}~\cite{koed-12}. \hl{This is in contrast to many other approaches in the
learning sciences that focus on the institutions that create learning 
opportunities, instruction styles, the roles of facilitators in learning
environments, and taxonomies of learning outcomes}~\cite{means2014learning}.
\section{Results}

Overall, we find that 93\% of our 2260 survey respondents reported learning 
something online. In this section, we detail the results of our three research 
questions, examining: (RQ1) what subjects people learn about online; (RQ2) how 
people learn online; and (RQ3) which online learning experiences are shared 
among the majority of online learners.

\subsection{What do people learn about online?}

\begin{figure}[htb]
\includegraphics[width=1\textwidth]{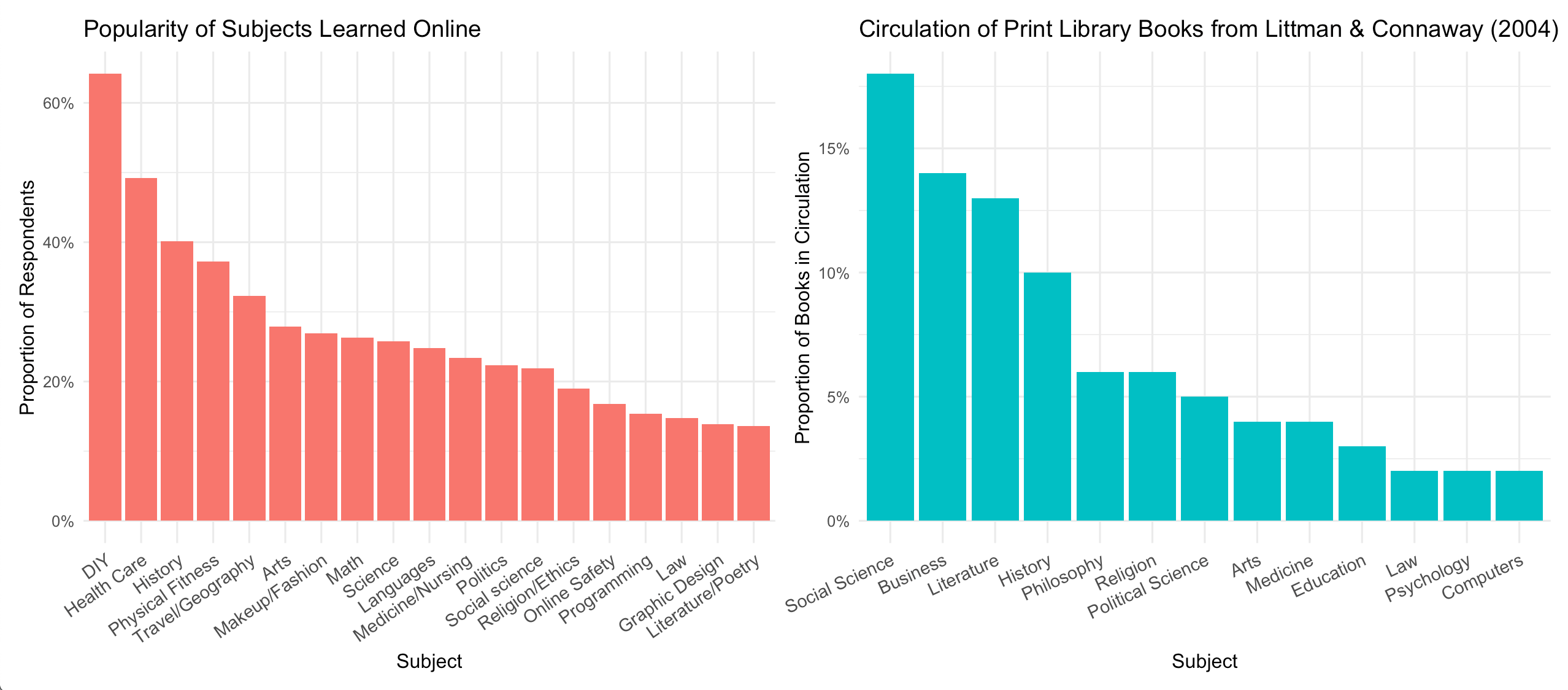}
\vspace{-3ex}
\caption{(Left) The popularity of different subjects that respondents learned online.
(Right) The proportion of books in circulation according to a study by Littman and Connaway ~\cite{Littman-2004}. Both the internet and library collections are public repositories for learning resources on many different subjects. This comparison shows that the distribution of subject popularity in our survey mirrors the same power law distribution found in the circulation of different subjects of library books.}
\label{fig1}
\vspace{-1ex}
\end{figure}

We asked survey respondents to report what subjects they learned about online. 
They reported learning about 19 subjects included in our question\footnote{As discussed in more detail 
in the methodology section, we also offered a free-text ``Other'' option for 
inputting additional subjects. Given that less than 1\% of respondents reported a subject 
not included in the subject list, and there was very little overlap in the 
subjects reported by these respondents.}. These subjects ranged from 
non-academic topics such as travel and personal physical fitness to 
traditional academic subjects such as law and mathematics 
(see Figure ~\ref{fig1} [Left] for a summary). 

%
We examine how many of our respondents chose to learn about these subjects. The only subject that more than half of our respondents reported learning about was DIY (62.4\%), which includes around-the-house activities such as cooking, baking, and arts-and-crafts. After DIY, the next most popular subjects, which were statistically less popular (p < 0.001 with Bonferroni-Holm correction for multiple comparisons)~\footnote{Statistical comparisons for each of the subject categories reported in this section can be found in the appendix.}, were those about general welfare and lifestyle -- physical health and healthcare (49\%), personal physical fitness (37\%), and travel and geography (32\%) -- and history (40\%).

Our results show that nearly all general-interest subjects were more popular than academic-focused subjects.
Academic subjects, aside from history, were significantly less popular (p < 0.005 Bonferroni-Holm correction for multiple comparisons) than general interest subjects. Less than 30\% of respondents reported learning about: fine arts (28\%), mathematics (26\%), the natural sciences (26\%), languages (25\%), medicine (23\%), politics (22\%), social sciences (22\%), religion and ethics (19\%), computer programming (15\%), law (15\%), graphic design (14\%), and literature (14\%).

%
To contextualize our findings, we compare the popularity of subjects learned online with the popularity of subjects learned offline. We find that while the subjects adults learn online vs. offline differ, the popularity of learning various subjects online and offline follows a similar trend: a few subjects are very popular while each of the remaining subjects are learned by a sizable minority. This is illustrated by the comparison in Figure ~\ref{fig1}: on the left, we show the popularity of subjects for online learning in our data set while on the right we show the distribution of offline learning interests as represented by the distribution of library books currently in circulation according to the Library of Congress Classification~\cite{Lancaster-1982}. We observe that this distribution follows a power law relationship like the Zipf distribution, where only a few subjects are very popular and most subjects are all approximately equal in their middling popularity.


\subsection{How people learn online}

\begin{figure}[htb]
\includegraphics[width=0.8\textwidth]{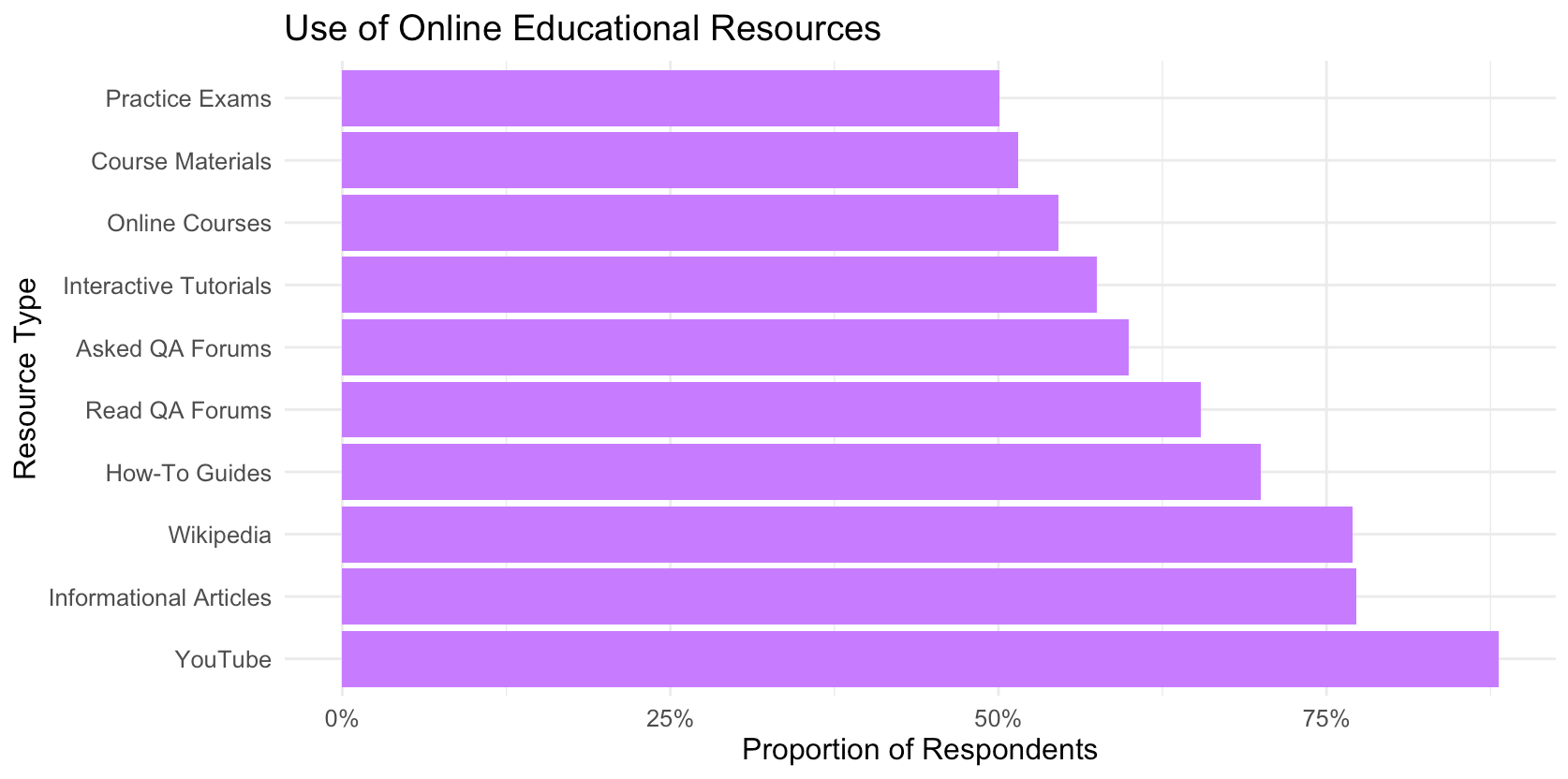}
\vspace{-3ex}
\caption{Proportion of respondents that reported using each type of educational resource.}
\label{fig:er}
\vspace{-3ex}
\end{figure}

Respondents used at least one and at most 10 (mean: 2.93, median: 2) resources to learn about a particular subject.

%
Of the 2094 respondents to our survey who indicated using at least one resource, the vast majority (88\%) relied on YouTube to learn online.\footnote{A respondent is considered to have learned online using a resource if they learned about at least one subject using this resource.}Additionally, 77\% used Wikipedia or informational articles, respectively, to learn online.
Over half of respondents reported learning online using how-to guides (70\%) or by using question-asking forums: either asking questions (60\%) or reading answers to others' questions (65\%). 
The least used online learning resources, but still used by at least half of respondents, were academic-style resources: interactive tutorials (58\%), course materials (e.g., slides, course notes) (52\%), and practice exam websites (50\%).

Further, 55\% of respondents reported enrolling in formal online courses to learn something. Of these respondents, 40.5\% reported having taken an online course from a local university, while 22.7\% said they had taken an online course at a university that was not local to them. The difference of these two proportions suggests that respondents' awareness of online education is not as global as the potential online educational opportunities that may be open to them. However, it is also possible that the higher percentage of locally-based online course enrollment may be influenced by college degree programs that combine in-person and online coursework, rather than by a lack of awareness about non-local options. 

Respondents also took online courses through professional services (20.5\%) such as Lynda, or through massively open online courses such as those offered on Coursera (16.5\%). A tenth (10.2\%) of respondents reported that they had used more than one type of online course. 

Figure ~\ref{fig:er} provides a summary of respondents' online learning resources.

\subsubsection{Exploring the relationship between how people learn and what they learn online}
Having identified the popularity of subjects and of resources, the next step is to look at the relationship between the two. Are certain resources used for more subjects than others? Are certain resources more popular for certain subjects than others?

\begin{figure}[htb]
\includegraphics[width=0.8\textwidth]{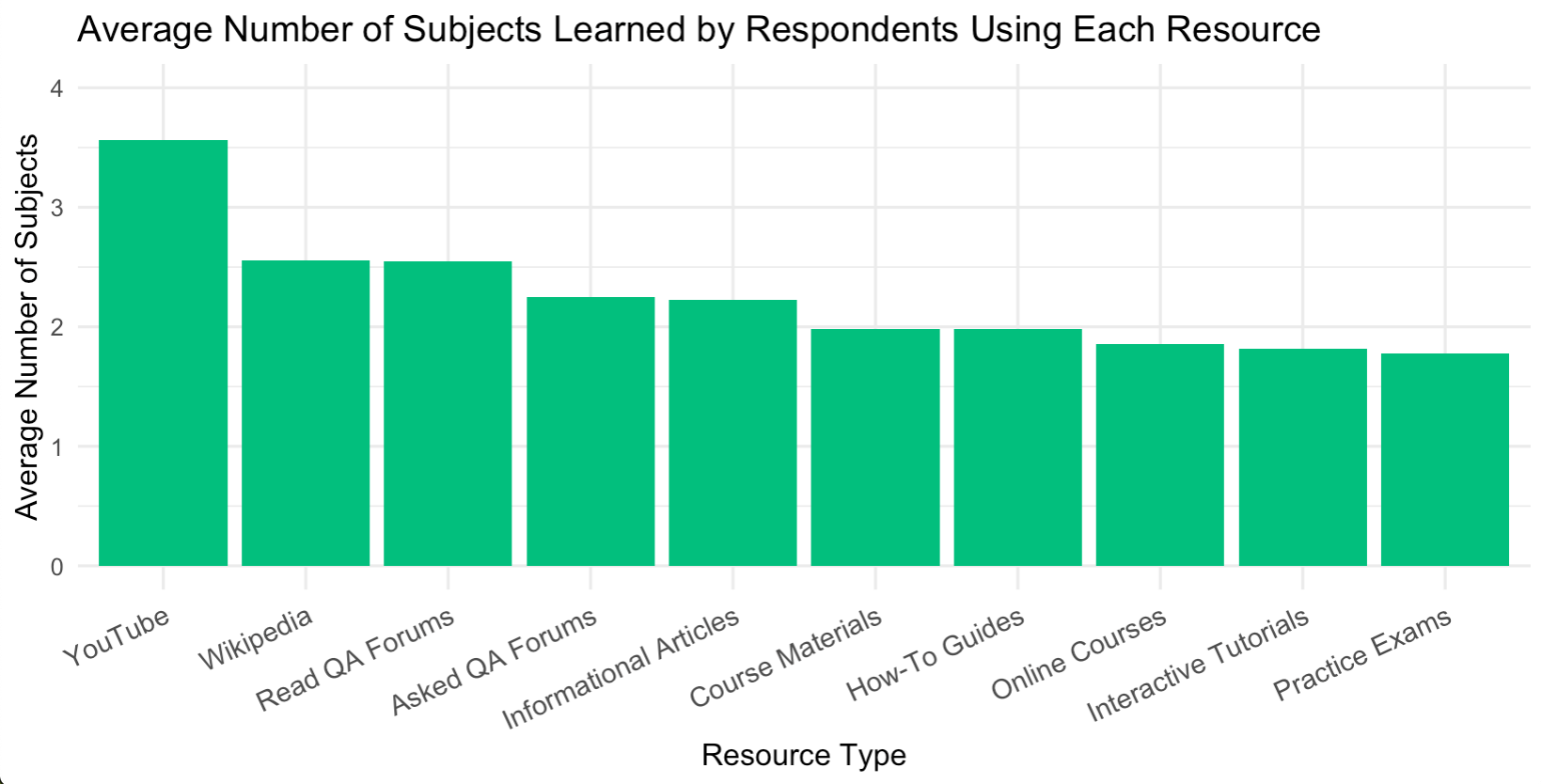}
\vspace{-3ex}
\caption{The average number of subjects that an individual respondent learned about for each resource.}
\label{fig:avgsub}
\vspace{0.1em}
\end{figure}

%
We find that YouTube is not only the resource used by the most respondents, it is also used by respondents to learn about the most subjects: a respondent who uses YouTube for learning uses the platform to learn about an average of 3.56 subjects. YouTube is used to learn about significantly more subjects (Mann-Whitney U Test, Bonferroni-Holm corrected p < 0.001) than Wikipedia (mean = 2.54 subjects) and other informational articles (mean = 2.25 subjects) or reading (mean = 2.54 subjects) and asking questions on Q\&A forums (mean = 2.25 subjects). All of these resources were used by respondents to learn significantly more subjects (Mann-Whitney U Test, Bonferroni-Holm corrected p < 0.002) than how-to guides, academic-style resources, and online courses, all of which were used to learn about less than two subjects, on average. Figure ~\ref{fig:avgsub} summarizes the average number of subjects that respondents learned about using each resource.

\begin{figure}[htb]
\includegraphics[width=1\textwidth]{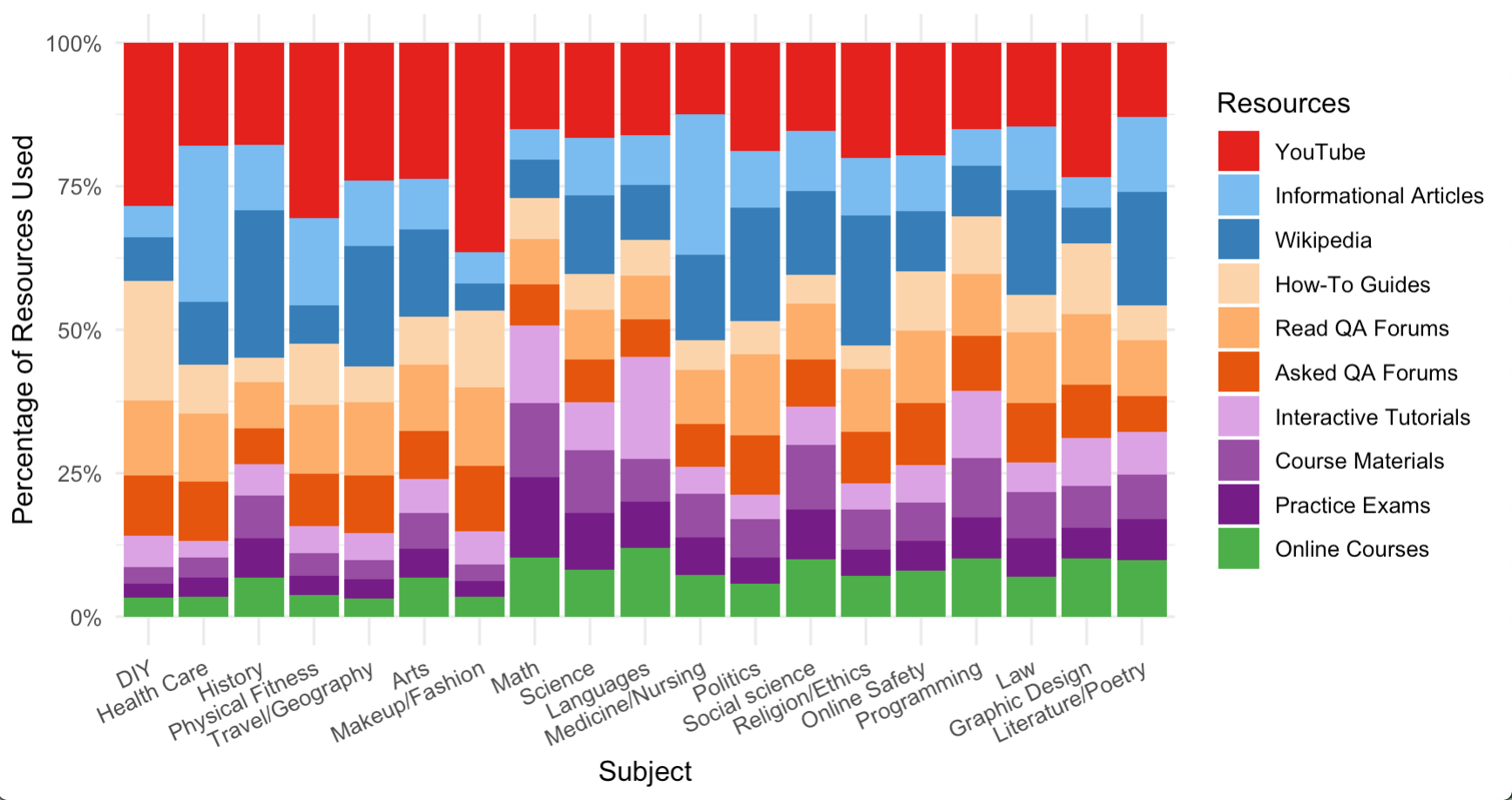}
\vspace{-3ex}
\caption{Proportion of each type of resource that was used to learn about each subject, in order from most popular to least popular subject.}
\label{fig:karyo}
\vspace{-3ex}
\end{figure}

%
While YouTube is the most used resource overall (used by 88\% of respondents), it is used particularly heavily to learn about certain subjects. Nearly a quarter of respondents who used YouTube learned about these five subjects: makeup and fashion (used by 36.5\% of learners), physical fitness (30.6\%), DIY (28.4\%), travel/geography (24.0\%) and the arts (23.7\%). Across all subjects, YouTube was at least the third most used type of learning resource. 

%
The next most popular resource types include informational articles and Wikipedia (used by 77\% of learners, respectively). Informational articles were particularly popular for learning about subjects related to health, including being the most popular resource for personal health care (27.2\%), medicine and nursing (24.6\%), and the second most popular resource for physical fitness (15.2\%). Wikipedia on the other hand was the most popular resource for learning social science and humanities subjects such as history (25.7\%), religion and ethics (22.7\%), literature and poetry (19.8\%), politics (19.7\%), and law (18.3\%). 

%
While how-to guides are, unsurprisingly, used by 20.7\% of respondents to learn about DIY, they are not otherwise used heavily to learn about any other particular subject (an average of 8.2\% of participants used how-to guides to learn across all subjects, SD = 4.04\%). Similarly, reading and asking questions on Q\&A forums made up 11.0\% (SD = 2.04\%) and 8.9\% (SD = 1.64\%), respectively, of the resources used by respondents to learn about any given subject.

%
Together, interactive online tutorials, course materials, and practice exams made up 40.5\% of the resources that respondents used to learn about Math. Similarly, these academic-style resources made up 33.3\% of the resources used to learn languages, 29.0\% of those used to learn programming, and 26.7\% and 29.3\%, respectively, of the resources used to learn social and natural science. The specific type of academic-style resource used most heavily to learn these subjects differed, with practice exams (14.0\% of the resources used) and interactive tutorials (13.6\% of the resources used) being particularly popular for math, interactive tutorials being particularly popular for languages (17.8\% of the resources used), and course materials being particularly popular for the social and natural sciences (~11\% of the resources used, respectively) and programming (10.2\% of the resources used).

\begin{table}[ht]
\centering
\begin{tabular}{lrrll}
  \hline
  Subject & Proportion Formal & Proportion Informal & Significance & Greater Proportion \\ 
  \hline
 History & 0.080 & 0.078 &  &  \\ 
   Arts & 0.056 & 0.055 &  &  \\ 
   Medicine/Nursing & 0.049 & 0.045 &  &  \\ 
   Online Safety & 0.039 & 0.032 &  &  \\ 
   Religion/Ethics & 0.035 & 0.032 &  &  \\ 
   Politics & 0.034 & 0.040 &  &  \\ 
   Law & 0.026 & 0.025 &  &  \\ 
   Math & 0.101 & 0.063 & *** & Formal \\ 
   Languages & 0.086 & 0.045 & *** & Formal \\ 
   Science & 0.076 & 0.061 & * & Formal \\ 
   Social science & 0.075 & 0.048 & *** & Formal \\ 
   Programming & 0.058 & 0.036 & *** & Formal \\ 
   Graphic Design & 0.042 & 0.026 & *** & Formal \\ 
   Literature/Poetry & 0.037 & 0.024 & ** & Formal \\ 
   DIY & 0.063 & 0.132 & *** & Informal \\ 
   Health Care & 0.047 & 0.091 & *** & Informal \\ 
   Physical Fitness & 0.035 & 0.063 & *** & Informal \\ 
   Travel/Geography & 0.024 & 0.052 & *** & Informal \\ 
   Makeup/Fashion & 0.022 & 0.043 & *** & Informal \\ 
   \hline
\end{tabular}
\caption{Pairwise statistical tests comparing the proportion of respondents who used formal versus informal online learning resources to pursue each subject. Zero * indicates no significant difference, one * indicates p < 0.05, two ** indicates p < 0.01, three *** indicates p < 0.001. `Greater Proportion' indicates whether formal or informal resources were used significantly more often.}
\label{tab:pair}
\end{table}

%
As mentioned in the introduction of this paper, the vast majority of prior work on online learning focuses on formal online courses rather than informal resources. This paper is the first, to our knowledge, to consider both types of resources. As such, we were especially interested in the differences between how people use online courses vs. informal online learning resources. These differences are summarized in Table ~\ref{tab:pair}.
We find that STEM-related academic subjects as well as social science, literature, languages, and graphic design are more often learned about using online courses than informal resources. 

On the other hand, informal resources were significantly more likely to be used to learn about general interest subjects: DIY, health care, physical fitness, travel and geography, and makeup and fashion. The contrast between these two groups of subjects may be related to the need (or lack of need) for a formal environment to become proficient in a subject. Mastering a language requires an experience approximating immersion into a new social environment, and learning graphic design may require hours of orientation with complex user interfaces. In contrast, successfully planning a trip or learning how to repair something specific around the house does not require intense prolonged study. Another difference between these two groups is the extent to which these activities are learned through physical interactions. Mathematics and computer programming are subjects concerned with manipulating abstract symbols, while training one's physical fitness or learning a new sewing or makeup technique have to be physically practiced to be understood. 

Finally, respondents in our survey were equally as likely to report having used an online course or an informal resource to learn about: history, arts, religion and ethics, politics, law, medicine and nursing, and online safety. We hypothesize that this is because these subjects straddle the line between academic and general interest: for example, an online learner could be studying art history as part of a degree program or job preparation or an online learner could be passively curious about the art of Michelangelo after a recent trip to Rome.

\subsection{Identifying core online learning experiences}

\begin{figure}
\includegraphics[width=0.8\textwidth]{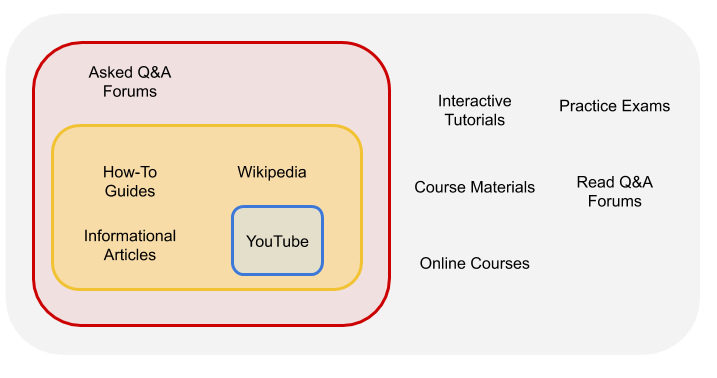}
\vspace{-3ex}
\caption{Euler diagram representing k-modes clustering of types of online learning resources. Our clustering analysis revealed three overlapping core sets of online learning resources. Resources that are contained within a greater number of circles are more central overall to online learning experiences. For example, YouTube is contained within three circles, compared to Wikipedia which is only contained within two circles. Therefore YouTube is a more central resource compared to Wikipedia.}
\label{fig:cr}
\vspace{-3ex}
\end{figure}

Finally, going beyond the popularity of what and \emph{how} people learn online, we consider common patterns in people's practice of online learning. We seek to identify shared experiences in online learners' interests (what they learn), learning tools (how they learn), and overall online learning experiences (common pairs of subjects and resources). 

First, to identify commonalities in online learners' interests, we used k-modes clustering to group respondents who reported learning more than one thing online based on the subjects they reported learning.
Of those who learned at least two things online, 76.1\% of respondents had in common only that they learned about DIY subjects online. These respondents did not necessarily indicate that they learned exclusively about DIY subjects, however their choice to learn about DIY subjects is what they had most in common.
An additional 18.8\% of respondents had in common that they learned both about DIY \emph{and} about at least one of: languages, health care, math, history, natural science, social science, and physical fitness. For the remaining 5.1\% of respondents, our clustering was unable to identify a common pattern among these respondents' online learning interests.

Second, we consider commonalities in how people learn online. We again used k-modes clustering to group those respondents who used more than one online learning resource, this time clustering the respondents based on the resources they used rather than the subjects they learned. 

Of respondents who used at least two online learning resources, 17.1\% had in common only that they used YouTube (the blue cluster in Figure ~\ref{fig:cr}), while 12.9\% of respondents had in common that they used both YouTube and at least one of the following: how-to guides, Wikipedia and informational articles. An additional 18.8\% of respondents had in common that they (1) used YouTube, (2) used at least one of the aforementioned three resources, and (3) asked questions in Q\&A forums. We could not find any pattern in resource use for the remaining 51.2\% of respondents. 

%

%


%
Of the 190 online learning experiences respondents reported, these twelve were experienced most often by our respondents: using YouTube to learn about DIY subjects (54.9\% of respondents), health care (24.0\%), makeup/fashion (23.2\%), history (21.2\%), arts (19.9\%), or travel/geography (18.3\%); using how-to guides to learn about DIY subjects (40.0\%), using informational articles to learn about health care (36.5\%), using Wikipedia to learn about history (30.6\%), or reading Q\&A forums or asking questions on those forums to learn about DIY subjects (25.4\% and 20.3\%, respectively). 


\section{Discussion} 
\label{sec:discussion}

Overall, we find that online learning is a ubiquitous experience: 93\% of 
respondents in our survey reported learning online. YouTube, in particular, is by far 
the most popular online learning resource. Not only is YouTube used by 
the most respondents for learning (88\%), it is the most popular resource 
for learning about 11 out of the 19 subjects examined in our survey. The 
fact that, on average, respondents used YouTube to learn about more than 
three subjects further underscores the importance of this resource. Moreover, 
YouTube is not only popular, it is core to the experience of 
online learning. Our clustering analyses consistently show that 
YouTube is at the core of other online learning experiences. This result is consistent with previous studies that have found that 
YouTube spans across formal and informal online learning resources~\cite{Rosenthal-2018}. YouTube recently redesigned an influential page where trending videos are displayed, adding a ``Learning'' category~\cite{yt-edu}, perhaps in response to this trend.  

In this section, we discuss 
which mechanisms of online learning are most pervasive and what the 
prevalence of particular mechanisms implies about the online learning 
ecosystem and how we design for it.
Additionally, we point out directions for future work based on our results.
We conclude with a look toward the 
future: what implications our findings have for the future of online 
learning in a world after COVID-19.

\subsection{Supply, Familiarity, and Ease As Potential Drivers of Leaning Resource Choice}

As noted earlier, YouTube is the most popular resource for online learning. Perhaps 
unsurprisingly, informational articles and Wikipedia are the next most 
popular types of online learning resources considering that text-based 
articles and web pages were the first and most fundamental parts of the 
internet ~\cite{tbl-www}. Indeed, the four core resources identified by our clustering -- YouTube, 
informational articles, Wikipedia, and how-to guides -- are built around 
three legacy internet technologies: video, static text, and wikis. YouTube 
was founded over 15 years ago ~\cite{yt-old}, static text websites have 
been around since the beginning of the internet, and wikis are a technology 
that is over 25 years old ~\cite{ebersbach-2008}. 

We hypothesize that the popularity of these legacy resources has four potential causes. First, adult learners may find it easiest -- or prefer -- to learn through informal video media (e.g., YouTube) and informational articles, and thus leverage these resources the most. Second, due in part to having a long period of time over which to develop content, these popular learning resources are able to supply a high volume of content across many topics. For example, YouTube is the leading video platform -- in terms 
of both hours of content and revenue~\cite{yt-content,yt-money}. As such, it may be easiest for learners to find learning content on YouTube, thus leading them to learn using video media. Third, over extended use, Internet users' may have developed \textit{familiarity} with learning from video and/or informational-article style content. This familiarity may lead them to continue to turn to learning modes from which they are comfortable receiving content instead of exploring newer learning technologies such as interactive tutorials, MOOCs and question-and-answer forums, which may also have a lower supply of content. Finally, the popularity of these resources may create a cyclical effect: because these resources are more popular, and offer more supply, they are easier to find in search results, and thus learners are more likely to turn to them. For example, prior work has suggested that some of 
Wikipedia's popularity could be attributed to how highly it is ranked in search engine results ~\cite{mcmahon-2017}. These findings raise important questions for future work to investigate regarding whether these resources actually best meet learner needs, or are merely used out of the convenience due to supply, familiarity, and/or ease of access. We delve deeper into how our findings support or refute each of these potential drivers behind learners' choice of online learning resources below.

\subsection{Low Subject Diversity Among Formal Resources}
%
Our work finds that adult online learners
have extremely diverse interests, \hl{with more than half (54\%)} learning about 5 or more different subjects. As mentioned above, one of the advantages of 
YouTube and Wikipedia (as well as informational articles) is supply: 
these resources host content about a wide array of subjects. However, newer 
online learning resources may lack this volume of content, either by choice or due to 
their relative novelty. This lack of diversity may inherently influence the mechanism 
of online learning. Let us take as a case study law and programming. 

Law and computer programming carry wide-ranging societal influence but were 
two of the least popular subjects among our respondents, with about 15\% of 
respondents indicating they learned about either subject. Despite the equal 
popularity of these subjects, programming was one of the most popular 
subjects for two types of resources (course materials, interactive tutorials), 
while law was 12th most popular of the 19 subjects respondents learned 
using Wikipedia, and ranked lower in terms of popularity for every other 
type of online learning resource. While it is possible that respondents 
use a wide variety of resources for learning about law because of the 
nature of the subject, this may also be the case because less online 
course content is available for law vs. for programming. Programming 
has numerous free, online course style resources such as Codecademy and open source 
scholarly projects like Python Tutor ~\cite{sims-2011,guo-2013} that put 
interactive programming tutorials online, in addition to more traditional 
online courses -- either free or payment-based -- such as those offered by universities or through 
MOOCs. Our findings suggest that learners may turn to newer resources tailored to their learning needs for a particular subject if those resources are available. \hl{In the absence of such tailored resources, however,} online learners may turn to the resources that they know will offer sufficient, affordable, supply: YouTube and Wikipedia.

\hl{By highlighting this contrast between law and computer science we do not mean 
to suggest that low subject diversity among formal online learning resources is
necessarily a problem to be addressed. We believe that this example illustrates 
with data that the way people learn different subjects online is nuanced in ways
that are not intuitive. Understanding these kinds of gradations in the online
learning landscape may also call attention to opportunities for new kinds of 
online learning experiences.}


\subsection{Issues of Supply May Relate to Ease of Resource Creation}
These issues of supply may relate to the ease with which resource creators can develop new content. 
%
 The relatively low popularity of new learning resources (e.g., MOOCS, interactive tutorials) highlights the absence of the diffusion of 
 innovations ~\cite{diffusion-1985} we might expect given years of interest 
 in new types of educational media and interactive online learning 
 technologies ~\cite{hazen2012proposed}. This absence may be related to the ease with which these resources can be 
 created. Phones, tablets, and personal computers are now often embedded with more 
 than one camera, making the creation of (educational) videos easier than 
 ever before, and the tools for publishing text-based articles are just as 
 widespread. However, relatively less robust and cost-effective support is available for the creation of newer resources. For example, many universities have had to scale up production studios in which to create MOOCS~\cite{baker2016value}, purchasing expensive equipment out of reach for many of the learning content creators who utilize YouTube. Similarly, many interactive tutorials involve the creation of new software products, requiring significant grant funding and time investment for development~\cite{sims-2011,guo-2013}. Thus, future work on the democratization of online education~\cite{kross2020democratization} may wish to consider how to democritize the \textit{creation} of content, in an effort to improve the learning content available for consumption. 

\subsection{Future work: Understanding Learners' Motivations}
Our results illustrate the breadth of adult online learning and raise a number of questions regarding why adults choose to learn using particular resources. Above, we hypothesized that learners may choose to learn using certain resources due to the supply of content available from certain resources, learners' familiarity with the mode of content, the ease with which learners can find that content, and with which educators can create it. However, online learners may also choose to learn particular topics with particular resources for reasons related to their own internal motivations, rather than externalities of the online learning ecosystem.
%
%
%

As discussed in prior work, there exists a spectrum of learning methods that may be used by, or which appeal to, different learners. This spectrum ranges from free choice learning -- e.g., visiting a museum, watching a documentary for entertainment~\cite{falk-2007} -- to traditional, academic, structured learning (e.g., taking a course). One of the contributions of our work is to answer the call~\cite{schwier-2012} to compare online learners' use of such informal vs. formal online resources. We find significant differences in the use of such resources (Table~\ref{tab:pair}). 

We hypothesize that these differences may relate not only to supply, but to learners' goals. For example, credentials earned online are increasingly popular
across many fields, and demand for these credentials only seems to be growing
~\cite{caudill2017emerging}. We hypothesize that learners who want to earn a credential 
are using online courses, practice exams, and course materials more often
compared to other online learning resources. In Figure \ref{fig:avgsub} we
can see that some of the subjects that most commonly use those resources are
math, programming, science, and languages, all subjects in which it can be
valuable to obtain credentials for professional purposes as opposed to the topics that we find are learned more frequently using informal resources such DIY and personal health care. Thus, we encourage future work building on the foundation laid by these findings, to investigate how the professional and credential-related motivations of online learners inform their choice of learning resources.

In addition to professionally-related credentials, our findings also suggest that learners may be motivated to seek more authoritative information about some topics, even if they are not pursuing a professional qualification. For example, respondents indicated that they used Wikipedia most for learning 
about history, politics, religion \& ethics, and law (see Figure
\ref{fig:avgsub}), all topics that may be presented with significant bias 
elsewhere. The Wikipedia community actively moderates their articles, lending a sense of \textit{community authority} ~\cite{sahut2017wikipedia} to their content, which may appeal to certain learners, or learners of particular subjects. As such, future work may seek not only to investigate learners' goals, but also the criteria through which they evaluate potential sources of learning information -- much like prior work has studied news consumers' evaluation of media and misinformation~\cite{scheufele2019science,lazer2018science}.

\subsection{The Rapidly Changing Online Learning Landscape}

During the preparation of this manuscript, COVID-19 emerged causing major 
disruptions and reorganization to how educational experiences are structured 
and delivered. Major universities had to transition all of their courses online
quickly to minimize in-person interactions and to comply with new government 
regulations designed to protect public health ~\cite{ed-corona}. We believe 
it is reasonable to say that the emergence of COVID-19 is the most significant 
event in the nascent history of online education, as the pandemic precipitated 
a situation where education in any form became nearly synonymous with online 
education.

Although the long-term effects of COVID-19 on education at large have yet to 
be realized, several short-term effects related to this study are taking form. 
Since the proliferation of the disease and the resulting lockdowns, more 
adult learners have been seeking online educational 
experiences ~\cite{adult-learners-online}. This may be related to the 
financial recession caused by COVID-19, since it is well understood that 
recessions drive increased interest in adult educational 
programs ~\cite{economy-school}. However, given that most college 
campuses are closed, adult learners are more frequently turning to online 
learning resources ~\cite{adult-learners-online}. Massively open online 
course providers in particular have seen a record-breaking growth 
in enrollments ~\cite{remember-moocs}. This is a significant development 
in the trajectory of MOOCs as prior work has reflected on the role of MOOCs 
in online education as modeled by their path through the Gartner Hype 
Cycle ~\cite{hype-cycle,kross-2018}.

The data in this study represent a snapshot of the adult online learning 
landscape just months before COVID-19. We believe that our results hold 
valuable insights that can inform the future design of online learning 
technologies. However, this study also inadvertently and advantageously 
provides a baseline that can be used in the future to measure how 
this monumental shift in online learning is changing interests in 
subjects or the differential use of types of online learning resources as 
both are influenced by the global pandemic. 

\section{Conclusion} 

In this work, we presented the results from a survey of 2260 adults about their
experiences using online learning resources to learn about different topics. Our
survey was balanced to include a diverse sample.
We found that over 90\% of respondents reported learning online about subjects
as diverse as math, politics, and the arts. The most popular subjects that
respondents learned about online included DIY, personal health care, history,
physical fitness, and travel \& geography. The proportion of respondents
interested in the various subjects in our survey followed a power law distribution.

The most popular types of online learning resources included YouTube, informational
articles, and Wikipedia. On average, participants used YouTube to learn about three
different subjects, and Wikipedia and informational articles to learn about
more than two different subjects. Respondents reported using every resource in
our survey to learn about every subject. The most popular subject on YouTube
was makeup \& fashion, while the most popular subject on Wikipedia was history,
and personal health care for informational articles.

Further analysis of our results showed that respondents learned about subjects
like math, law, and computer programming from more formal online resources. However,
for subjects like DIY, physical fitness, and travel \& geography, respondents
more often learned from informal resources. Some respondents used formal and
informal online learning resources equally to learn about subjects like history,
the arts, and politics. Finally, a clustering analysis revealed that YouTube is 
the resource that has the most overlap among all of our respondents' online 
learning experiences.

To conclude, we discussed the importance of video and text as technologies that
are fundamental to the web itself, and serve as especially important 
infrastructure for online education. We also explored how the forces of supply
and demand drive the availability of online educational experiences. Finally,
we set our sights on the future by examining factors that impact resources
that learners seek out, and how online education will be shaped by the COVID-19
pandemic.

\begin{acks}
We would like to thank Facebook Research for funding this study as part of
their Economic Opportunity and Digital Platforms Research Award. We would also
like to thank Tamara Clegg and Heather Killen for their helpful input.
\end{acks}

\bibliographystyle{ACM-Reference-Format}
\bibliography{cited}

\received{June 2020}
\received[revised]{October 2020}
\received[accepted]{December 2020}

\end{document}